\documentstyle[aps,preprint]{revtex}
\draft

\begin{document}
\title
{A note on Coulomb scattering amplitude}
\author{Zafar Ahmed}
\address{Nuclear Physics Division, Bhabha Atomic Research Centre,
Bombay 400 085\\
e-mail:zahmed@apsara.barc.ernet.in}
\date{\today}
\maketitle
\begin{abstract}
The summation of the partial wave series for Coulomb
scattering amplitude, $f^C(\theta)$ is avoided because the series is
oscillatorily and divergent. Instead, $f^C(\theta)$ is obtained by solving
the Schr{\"o}dinger equation in parabolic cylindrical co-ordinates which is
not a general  method. Here, we show that a reconstructed series, $(1-\cos\theta)
^2f^C(\theta)$, is both convergent and  analytically summable.
\end{abstract}
\vskip 1 in
\par The partial wave analysis is the most widely  used method to obtain
cross-sections of the scattering processes due to central potentials in atomic,
nuclear and molecular physics [1]. In this method the scattering matrix $S_l$
is the most important theoretical ingredient which is the characteristic of a
scattering potential. By knowing $S_l$ one can obtain the scattering amplitude,
$f(\theta)$, and hence the cross-section , $\sigma (\theta)(=|f(\theta)|^2).$
The partial wave series for the scattering amplitude which sums over discrete
angular momentum, $l$, is written in terms of $S_l$ as [1]
\begin{equation}
f(\theta)={1 \over 2ik} \sum_{l=0} ^{\infty} (2l+1) [S_l-1] P_l(\cos\theta).
\end{equation}
Despite the importance of an S-matrix it is ironical that the Coulomb
potential, $V^C(r)={ Q_1 Q_2 \over r}$,
is  unique example admitting  a simple, exact and analytic S-matrix
for all angular momenta and energies. The Coulomb S-matrix is expressed as [1]
\begin{equation}
S^C_l=\exp[2i\sigma_l]={\Gamma(l+1+i\eta) \over \Gamma(l+1-i\eta)},
\end{equation}
where the super-script $C$ stands for Coulomb, $\sigma_l$ is the Coulomb
phase-shift and $\eta(={Q_1 Q_2 \mu \over {\hbar}^2 k})$ is the
Sommerfeld parameter, $\Gamma(z)$ is the gamma function. Surprisingly,
in books [1] the summation or summability of
the series (1) for the Coulomb potential remains  not only elusive
but also without a remark. Generally, the next thing that is done [1]
is to take
resort to solving Schr{"\o}dinger equation in parabolic cylindrical co-ordinates
to extract $f^C(\theta)$ as [1]
\begin{equation}
f^C(\theta)=-{\eta \over 2k\sin^2(\theta/2)} \exp[-i\{\eta \log (\sin^2(\theta/2))
-2\sigma_0\}]
\end{equation}
\par In this situation, one may wonder if the partial wave analysis is not a
general method for obtaining the scattering amplitude for a central potential.
Furthermore, if one tries to sum the series (1) inserting (2),
the exercise turns out to be rather frustrating in that the series oscillates
showing no convergence, irrespective of number of partial waves included [2,3].
Hence, one fails to reproduce $f^C(\theta)$ (3) even numerically. In fact Eq.(3)
serves as the benchmark result in the theory of scattering. For the charged
particle scattering e.g., proton scattering from nucleus where in addition
to the Coulomb potential, a short ranged  nuclear potential, $V_N(r)$, is used to
calculate the scattering amplitude,
\begin{equation}
f(\theta)={1 \over 2ik} \sum_{l=0} ^{\infty} (2l+1) [\exp(2i\eta_l)-1]
P_l(\cos\theta),
\end{equation}
where $\eta_l$ is the phase shift of the combined nuclear and Coulomb potential.
Once again, the series (4) would show oscillatory divergence and in order to
suppress this divergence one writes $\eta_l=\delta_l+\sigma_l$. Notice that
$\delta_l$ is the $\sigma_l$-subtracted phase shift which will also
depend on the parameter, $\eta $, in some way which, in turn, is not known
explicitly.
If $\eta_l$ corresponds to $V_N(r)+V_C(r)$ and $\sigma_l$ corresponds
to $V_C(r)$, one can really not assert as to which potential form in
terms of $V_C(r)$ and $V_N(r)$, the phase shift $\delta_l$ would actually
and exactly correspond to? Only in an approximate calculation such as Born
approximation [1] the phase shifts add for two potentials.
In any case Eq.(4) can be algebraically split as [1]
\begin {mathletters}
\begin{equation}
f(\theta)={1 \over 2ik} \left (\sum_{l=0} ^{\infty} (2l+1) [\exp(2i\sigma_l)-1]
P_l(\cos\theta)
+ \sum_{l=0} ^{\infty} (2l+1) \exp(2i\sigma_l) [\exp(2i\delta_l)-1]
P_l(\cos\theta) \right ).
\end{equation}
Subsequently, in the above expression the first series, instead of summing, it
is ingeniously substituted by $f^C(\theta)$ (3) as
\begin{equation}
f(\theta)=f^C(\theta)+{1 \over 2ik} \sum_{l=0} ^{\infty} (2l+1) \exp(2i\sigma_l)
[\exp(2i\delta_l)-1] P_l(\cos\theta).
\end{equation}
\end{mathletters}
By doing so,
the convergence of $f(\theta)$ is solely controlled by the $l$- dependence of
$[\exp(2i\delta_l)-1]$ which generally vanishes for large $l$. In the
standard text-books [1] Eq. (5b) is essentially written; however, the
motivation for doing so is not mentioned. It must be emphasized here that
the Eq. (5b) does not serve any purpose other than to disentangle the
divergence of (4, 5a). Equation (5b) is mistaken to separate out the effect of
Coulomb interaction from $f(\theta)$, in fact some effect of the Coulomb part
goes implicitly in the second part.
\par Alternatively, Yennie et al., [4] proposed to bypass this split-up
(5b) of $f(\theta)$ for charged particle scattering or for any other instance of scattering
where the series for $f(\theta)$ diverges. They suggested an intuitive numerical
recipe to calculate $f(\theta)$ from Eq.(4) itself. According to them,
one should rather reconstruct a reduced series i.e., $(1-\cos\theta)^m f(\theta)$
by choosing a positive integer, $m$, such that the resulting series shows
convergence. They have successfully calculated cross-sections of high energy
electron scattering by nuclei. Somehow, this prescription has not received as
much attention as it deserves.
\par Utilizing this prescription [4] we have been curious to know whether we
can overcome the divergence problem of Eq.(1) with Eq.(2) and extract the
Coulomb scattering amplitude (3) directly therefrom. We find that the
reconstructed series $(1-\cos\theta)^2f^C(\theta)$ is uniformly and absolutely
convergent and can be summed analytically to recover (3). It may be worthwhile
to mention that several elegant and rigorous methods of the summation of the
partial wave amplitudes
for long range potentials including the Coulomb potential have been proposed [7].
These works are of more general nature  which are numerical and which dwell
more upon approximating the Legendre polynomial $P_l(\cos\theta)$ in various
elegant ways.
\par The divergence of (1) for $S^C_l$ can be at once realized by noticing
the large $l$ limit of $S^C_l$. Using an asymptotic property of the
gamma function i.e., $ {limit \atop {z\rightarrow \infty}} \Gamma(z+a)
\rightarrow \Gamma(z) \exp[a(\log z)],$  [5] to
find that ${limit \atop {l\rightarrow \infty}} S^C_l \rightarrow
(l+1)^{2i\eta} .$ It shows that the quantity
$[S^C_l-1]$ oscillates even for asymptotically large values of $l$.
\par Using a property of the Legendre polynomials [5,6] i.e., $\sum_{0}^{\infty}
(2l+1) P_l(\cos\theta_0)P_l(\cos\theta)=2\delta(\cos\theta_0-\cos\theta),$ we
can write
\begin{equation}
f^C(\theta)=(2ik)^{-1} \left( \sum_{0}^{\infty} (2l+1) S^C_l P_l(\cos\theta)
-2\delta(1-\cos\theta) \right ),
\end{equation}
where $\delta(z)$ is the Dirac-delta function. This shows an obvious
singularity
(divergence) in $f^C$ at $\theta=0.$ Let us see if we can suppress this
divergence. Recalling an interesting property of Dirac-delta function
i.e., $z\delta(z)=0,$ it is tempting to multiply (6) by $(1-\cos\theta$)
on both the sides. We then get rid of the second term and find
\begin{equation}
(1-\cos\theta)f^C(\theta)=\sum_{0}^{\infty} (2l+1) S^C_l (1-\cos\theta)
P_l(\cos\theta).
\end{equation}
Next we reconstruct this series by using a recurrence formula of Legendre
polynomials [5,6]
\begin{equation}
(2l+1)xP_l(x)=(l+1)P_{l+1}+ lP_{l-1}(x),
\end{equation}
where $\cos\theta=x$ and we get
\begin{equation}
(1-x)f^C(x)=(2ik)^{-1}\sum_{0}^{\infty}[ (2l+1)S^C_l P_l(x)-(l+1)S^C_l
P_{l+1}(x)-lS^C_lP_{l-1}(x)].
\end{equation}
\begin{equation}
(1-x)f^C(x)=(2ik)^{-1}\sum_{0}^{\infty} [(2l+1)S^C_l-lS^C_{l-1}-
(l+1)S^C_{l+1}] P_l(x),
\end{equation}
which in terms of gamma functions reads as
\begin{equation}
2ik (1-x) f^C(x)=\sum_0^{\infty} \left ((2l+1){\Gamma(l+1+i\eta)
\over \Gamma(l+1-i\eta)}
-l{\Gamma(l+i\eta) \over \Gamma(l-i\eta)}-(l+1){\Gamma(l+2+i\eta) \over
\Gamma(l+2-i\eta)} \right ) P_l(x).
\end{equation}
By multiple usage of the recurrence formula for the gamma function viz.,
$\Gamma(z+1)=z\Gamma(z)$, we can write
\begin{equation}
2ik(1-x)f^C(x)=\sum_{0}^{\infty} 2\eta^2(2l+1) \left ({\Gamma(l+i\eta) \over
\Gamma(l+2-i\eta)} \right ) P_l(x).
\end{equation}
For a use in the sequel, here let us denote the quantity appearing inside the
large bracket in the above equation  as $T_l$. The above series can be
re-written in terms of Coulomb phase-shifts as
\begin{equation}
2ik(1-x)f^C(x)=\sum_{0} ^{\infty} {2\eta^2(2l+1)\exp[2i\sigma_{l-1}]
\over (l-i\eta) (l+1-i\eta)} P_l(x).
\end{equation}
According to Weierstrass' M-Test [5] , a {\it sufficient} condition for a
series, $\{U_n(z)\}$, to be uniformly and absolutely convergent is that the
series, $\{M_n\}$, converges. Here, $M_n>|U_n(z)|$ for the given range of $z$,
where $M_n$ must be independent of $z$. Noting that $|P_l(x)|<1$ for
$-1\le x\le 1$, we choose
\begin{equation}
M_l={2\eta^2 (2l+1) \over \sqrt{(l^2+\eta^2)\{(l+1)^2+\eta^2\}}}.
\end{equation}
The series, $\{M_l\}$ upon comparing with $\sum_{0}^{\infty} 1/l$ (divergent)
is divergent. Hence, the M-test turns out to be negative for the uniform and
absolute convergence of the series (12,13). In fact, the M-test being only
{\it sufficient} we can not really be assertive about the convergence of (12,13).
It may be remarked that due to the presence of $P_l(x)$ in
Eq.(13) only M-test is feasible here. Therefore, we further reduce the series
(12) by multiplying by $(1-x)$ on both the sides in anticipation of a
better alternative series for which the M-test can be positive about its
convergence and hence about its summability. The second
reduced series with the help of Eqs. (8,12) is obtained as
\begin{equation}
2ik(1-x)^2f^C(x)= 2\eta^2 \sum_{0}^{\infty} [(2l+1)T_l P_l(x)-(l+1)T_l P_{l+1} -
l T_l P_{l-1}(x)].
\end{equation}
Carrying out similar manipulations as done
earlier from Eqs. (10) to (12), we reconstruct the series (15) as
\begin{equation}
2ik(1-x)^2f^C(x)=-4\eta^2(1-i\eta)^2 \sum_{0}^{\infty} (2l+1)
{\Gamma(l-1+i\eta) \over \Gamma(l+3-i\eta)} P_l(x).
\end{equation}
Eliminating gamma functions from this series, we get a more transparent
expression in terms of Coulomb phase-shifts.
\begin{equation}
2ik(1-x)^2f^C(x)=-4\eta^2(1-i\eta)^2 \sum_{0} ^{\infty} {(2l+1)
\exp[2i\sigma_{l-2}] \over (l+2-i\eta)(l+1-i\eta)(l-i\eta)(l-1-i\eta)}
P_l(x).
\end{equation}
Once again we carry out the M-test, choosing the M-series to be
\begin{equation}
M_l=\sum_{0}^{\infty} {(2l+1) \over \sqrt{ \{(l+2)^2+\eta^2\} \{(l+1)^2
+\eta^2\} \{l^2+\eta^2\} \{(l-1)^2+\eta^2\}}},
\end{equation}
which upon comparison with $\sum_{0}^{\infty} l^{-3}$ (convergent),
is convergent. This establishes the uniform and absolute convergence of (15,16).
Let us now sum it up. To this end, we make use of a very interesting formula
due to Bateman [6]
\begin{equation}
(1-x)^\rho=2^\rho \sum_{0}^{\infty} {2n+1 \over n+\rho+1}{(-\rho)_n
\over (1+\rho)_n} P_n(x),
\end{equation}
where $(\xi)_n=\Gamma(\xi+n)/\Gamma(\xi)$. By employing Eq.(18) with
$\rho=1-i\eta$, Eq.(16) yields
\begin{equation}
2ik(1-x)^2f^C(x)=-{4\eta^2(1-i\eta)^2 2^{-1+i\eta} (1-x)^{1-i\eta}
\Gamma(-1+i\eta) \over \Gamma(2-i\eta)}.
\end{equation}
By using the identity, $z\Gamma(z)=\Gamma(1+z)$ and putting $x=\cos\theta$,
we straightaway obtain the Coulomb  scattering amplitude (3). Thus, the
derivation of Eq. (3) presented here supplements the partial wave analysis
for the Coulomb potential. This exercise also demonstrates how the
divergence of a series is disentangled to extract a physical result.
\section *{References :}
\begin{enumerate}
\item e.g., L.I.Schiff, {\em Quantum Mechanics} (Mc Graw Hill, New York, 1968)
$3^{rd}$ ed. pp. 138-147;\\
E.Merzbacher, {\em Quantum Mechanics} ( John Wiley and Sons Inc.,
New York, 1970) $2^{nd}$ ed., pp. 245-250;\\
M.L.Goldberger and K.M.Watson, {\em  Collision Theory }
(John Wiley and Sons Inc., New York, 1964) pp. 259-269;\\
N.F.Mott and H.S.W. Massey, {\em The theory of Atomic Collisions} ( Oxford University Press, London, 1965)
$3^{rd}$ ed. , pp. 53-68.
\item L.Marquez, `` On the divergence of the Rutherford scattering amplitude in
terms of Coulomb phase shifts,'' Am. J. Phys. {\bf 40} 1420-1427 (1972).
\item J.R.Taylor,`` A new rigorous approach to Coulomb scattering,''
Nuovo-Cimento {\bf B 23} 313-334 (1974).
\item D.R.Yennie, D.G.Ravenhall and R.N.Wilson, `` Phase-shifts
calculation of high energy electron scattering,'' Phys. Rev. {\bf 95} 500-515 (1954).
\item E.T. Whittaker and G.N.Watson, {\em A Course in Modern Analysis},
(University Press, Cambridge, 1952) $4^{th}$ ed.
\item H.Bateman, {\em Higher Transcendental Functions,}
(McGraw-Hill, New York, 1953) vol. 2, p. 214.
\item e.g., A.K.Common and T.W.Stacey, `` The convergence of Legendre Pade' approximants
to the Coulomb and other scattering amplitudes,'' J.Phys.: Math. \& Gen. {\bf A 11}
275-289 (1978);\\
C.R.Garibotti and F.F.Grinstein, `` Summation of partial wave expansions in
the scattering by long range potentials II : Numerical applications,''
J. Math. Phys. {\bf 20}, 141-147 (1979);\\
F. Gesztesy and C.B. Lang, `` On the Abel summability of partial wave amplitudes for
Coulomb-type interactions,'' J. Math. Phys. {\bf 22} 312-319 (1981).
\end{enumerate}
\end{document}